 \documentclass[utf8]{FrontiersinHarvard} 

\usepackage{url,lineno,microtype,subcaption}
\usepackage[onehalfspacing]{setspace}
\usepackage{tikz}
\usepackage[colorlinks, linkcolor=blue, anchorcolor=blue, citecolor=blue, urlcolor=blue]{hyperref}
\DeclareRobustCommand{\orcidicon}{%
	\begin{tikzpicture}
	\draw[lime, fill=lime] (0,0)
	circle [radius=0.16]
	node[white] {{\fontfamily{qag}\selectfont \tiny ID}};
	\draw[white, fill=white] (-0.0625,0.095)
	circle [radius=0.007];
	\end{tikzpicture}
	\hspace{-2mm}
}
\foreach \x in {A, ..., Z}{%
	\expandafter\xdef\csname orcid\x\endcsname{\noexpand\href{https://orcid.org/\csname orcidauthor\x\endcsname}{\noexpand\orcidicon}}
}

\def\keyFont{\fontsize{8}{11}\helveticabold }
\def\firstAuthorLast{Y. T. Zhu {et~al.}} 
\def\Authors{Y. T. Zhu\orcidA{}\,$^{1,2,3}$, R. B. Wu\,$^{4}$, Z. H. Peng\,$^{5}$, and S. Xue\orcidB{}\,$^{1,2,3,*}$}
\def\Address{$^{1}$Department of Automation, Shanghai Jiao Tong University, Shanghai 200240, P. R. China\\
$^{2}$Key Laboratory of System Control and Information Processing, Ministry of Education of China, Shanghai 200240, P. R. China\\
$^{3}$Shanghai Engineering Research Center of Intelligent Control and Management, Shanghai 200240, P. R. China\\
$^{4}$Center for intelligent and networked systems, Department of Automation, Tsinghua University, Beijing 100084, P. R. China\\
$^{5}$Key Laboratory of Low-Dimensional Quantum Structures and Quantum Control of Ministry of Education, Key Laboratory for Matter Microstructure and Function of Hunan Province, Department of Physics and Synergetic Innovation Center for Quantum Effects and Applications, Hunan Normal University, Changsha 410081, P. R. China}
\def\corrAuthor{Shibei Xue}

\def\corrEmail{shbxue@sjtu.edu.cn}

\begin{document}
\onecolumn
\firstpage{1}

\title{Giant-cavity-based quantum sensors with enhanced performance} 

\author[\firstAuthorLast ]{\Authors} 
\address{} 
\correspondance{} 

\extraAuth{}

\maketitle

\begin{abstract}

\section{}
Recent progresses have revealed that quantum systems with multiple position-dependent couplings, e.g., giant atoms, can exhibit some unconventional phenomena, such as non-exponential decay etc. However, their potential applications are still open questions. In this paper, we propose a giant-cavity-based quantum sensor for the first time, whose performance can be greatly enhanced compared to traditional cavity-based sensors. In our proposal, two cavities couple to a dissipative reservoir at multiple points while they couple to a gain reservoir in a single-point way. To detecting a unknown parameter using this sensor, a waveguide is coupled to one of the cavities where detecting fields can pass through for homodyne detection. We find that multiple position-dependent couplings can induce an inherent non-reciprocal coupling between the cavities, which can enhance the performance of sensors. Output noise in our scheme can be reduced to the shot noise level, which is about one order magnitude lower than the results in Ref. [\cite{Lau2018}]. Besides, the signal-to-noise ratio per photon is also enhanced by about one order of magnitude. These results show that the multiple-point-coupling structure is beneficial to nowadays quantum devices. 

\tiny
 \keyFont{ \section{Keywords:} Giant Atoms, Giant Cavities, Quantum Sensors, SNR, and Non-Markovian} 
\end{abstract}

\section{Introduction}
High precision measurement of physical quantities lies in the core of metrology, e.g., gravitational wave detection [\cite{Schnabel2010, Stray2022}], nano-particles detection [\cite{Vollmer2008, Zhu2010, Zhao2011, Shi2014}], thermal sensing [\cite{Xu2018}], navigation [\cite{Sanchez-Burillo2012, Marks2014}], and magnetometers [\cite{Li2018A, Li2018B, Yu2016}]. Towards fundamental detection limits in weak-signal measurements, non-reciprocity [\cite{Sounas2017}] has become a powerful resource [\cite{Lau2018}]. Since reciprocity is hard to break due to Lorentz theorem [\cite{Potton2004}], many methods have been proposed for inducing non-reciprocity, for example, biasing with odd-symmetric quantities under time reversal [\cite{Casmir1945}], steering systems into exceptional points [\cite{Peng2014B, Peng2016}], constructing directional couplings [\cite{McDonald2020}], employing asymmetric or non-linear elements [\cite{Wang2013, Konotop2002, Scalora1994, Tocci1995, Zhukovsky2011}], or breaking the time-invariance of systems [\cite{Xu2005, Phare2015}]. 

 Recent progresses on quantum systems with multiple-point couplings (e.g., giant atoms  [\cite{Kockum2014, Schuetz2015, Manenti2017, Guo2017, Kockum2018, Andersson2019, Ekstrom2019, Delsing2019, Kannan2020, Kockum2020, Guo2020, Vadiraj2021, Du2021A, Du2021B, Du2021C, Cai2021, Zhu2021}]) provide a new possibility to acquire non-reciprocity. Due to the position-dependent relative phase between coupling points,  decays in the systems depend on the frequency of input fields [\cite{Guo2017, Zhu2021, Du2021C}], which provides a tunable and flexible approach to engineering atom-reservoir interaction. Besides, if several giant atoms are properly aligned to couple to a common reservoir, their coherence can be protected by the virtual interaction via this shared reservoir [\cite{Kockum2018}]. This unique phenomenon may lead to a new approach to acquiring non-reciprocity through the decoherence-free interaction depending on the input frequency,  although the multi-point-coupling structure itself cannot induce non-reciprocity in the absence of external non-trivial coupling phase [\cite{Du2021A}]. 

In this paper, we propose a quantum sensor consisting of two giant cavities, between which an inherent non-reciprocal coupling can be built up through a shared reservoir.  Comparing with microcavity-based structures [\cite{Lau2018}], the signal-to-noise ratio in our proposal can be improved by one order of  magnitude. The remaind of this paper is organized as follows. In Sec. 2, we propose the theoretical model of the quantum sensor, including the Hamiltonian and equations of motion. Following the standard frame [\cite{Lau2018}], we propose the performance indicator of sensors in Sec. 3, including signal, output noise, and signal-to-noise ratio per photon. The comparison with the sensor made up of small cavities is shown in Sec. 4. Finally, the further discussion and conclusion are given in Sec. 5.    

\section{Model of Gaint-Cavity-based Quantum Sensor }
\subsection{Hamiltonian}
\begin{figure}[h!]
\begin{center}
\includegraphics[width=14cm]{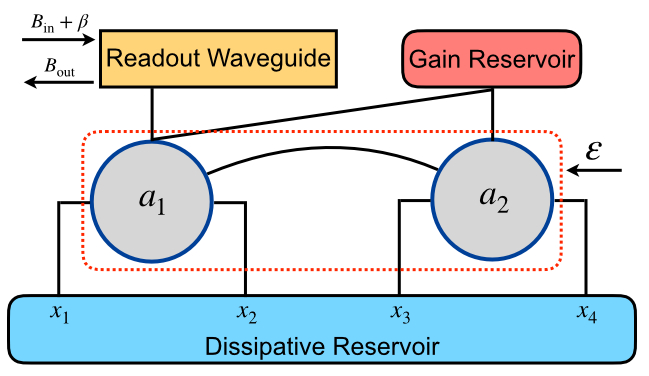}
\end{center}
\caption{Schematic of the two-giant-cavity quantum sensor. Both cavities couple to a dissipative reservoir at multiple points, i.e., $x_1$ and $x_2$ for cavity 1 (denoted by annihilation operator $a_1$), $x_3$ and $x_4$ for cavity 2 (denoted by operator $a_2$). The distance between the two points for one cavity is sufficiently large, which induces non-negligible time delays,  such that it forms two giant cavities. As a result,  the couplings between cavities and dissipative reservoir are position dependent. On the contrary, both cavities couple to the gain reservoir at the same point. Also, a classical drive $\beta$ with a noise input $B_{\mathrm{in}}$ is injected into the readout waveguide which only couples to the cavity 1. Its reflected field $B_{\mathrm{out}}$ is measured by homodyne detection. Initially, both reservoirs and the waveguide are prepared in the vacuum state. This sensor can reflect the external perturbation $\varepsilon$ from the variations of the output $B_{\mathrm{out}}$. }
\label{model}
\end{figure}
In Ref. [\cite{Lau2018}], a paradigm of quantum sensors is proposed that several coupled cavities couple to a gain reservoir and a dissipative reservoir at a single point. Illuminated by this paradigm, the sensor we considered consists of a coupled double-cavity  interacting with two reservoirs. The first cavity is coupled to a dissipative reservoir at $x_1$ and $x_2$,  and the second cavity is coupled to it at $x_3$ and $x_4$, as shown in Fig. \ref{model}. On the contrary, a gain reservoir couples to both cavities at the same point. Besides, a classical driving field $\beta$ with a noise input $B_{\mathrm{in}}$ enters the readout waveguide which only couples to the cavity 1, and its reflected field $B_{\mathrm{out}}$ is measured by homodyne detection. According to the model, the total Hamiltonian reads
\begin{equation}
H_{\mathrm{tot}}=H_0+H_{d}+H_I,	
\end{equation}
where
\begin{subequations}
\begin{eqnarray}
H_0=&&\sum_{\mathrm{i,j=1}}^2 H_{\mathrm{ij}}[\varepsilon]a^\dagger_\mathrm{i} a_\mathrm{j}+\int\mathrm{d}k \ \omega_{b,k}b_k^\dagger b_k+\int\mathrm{d}k\ \omega_{c,k} c_k^\dagger c_k+\int\mathrm{d}k\ \omega_{d,k} d_k^\dagger d_k,\label{Hamiltonian0}\\
H_d=&&\sqrt{\kappa}(\beta e^{-i\omega_L t}a_1^\dagger+H.c.)+\sqrt{\kappa}\int \frac{\mathrm{d}k}{\sqrt{2\pi}}\ \big(a_1^\dagger b_k+H.c.\big),\label{HamiltonianD}\\
H_I=&&\sum_{\mathrm{i}=1}^2 \int \mathrm{d}k\ \big(Y_\mathrm{i}a_\mathrm{i}^\dagger c_k^\dagger+H.c.\big)+\nonumber\\
&&\int \mathrm{d}k\ \left(Z_1(e^{ik x_1}+e^{ik x_2})a_1^\dagger d_k+Z_2(e^{ik x_3}+e^{ik x_4})a_2^\dagger d_k+H.c.\right).\label{HamiltonianI}
\end{eqnarray}
\end{subequations}
Eq. \eqref{Hamiltonian0} describes the free Hamiltonian of the two cavities, the readout waveguide, the gain and dissipative reservoirs with bosonic annihilation operators $a_{\mathrm{i}}$, $b_k$, $c_k$, and  $d_k$, respectively. Here, we have assumed that the perturbation $\varepsilon$ is small enough such that $H_{\mathrm{ij}}[\varepsilon]$ has a linear form \footnote{Since the perturbation is small enough, such that it can be expanded as a small quantity and kept to the first order.} [\cite{Lau2018, Bao2021}] $H_{\mathrm{ij}}[\varepsilon]=H^f_{\mathrm{ij}}+\varepsilon V_{\mathrm{ij}}$, where $H^f_{\mathrm{ij}}$ is the unperturbed part of the coupled cavities and $V_{\mathrm{ij}}$ denotes the coupling of perturbation $\varepsilon$ on the cavities. The first term in Eq. \eqref{HamiltonianD} represents a classical driving $\beta$ with a frequency $\omega_L$ and a coupling strength $\kappa$ enters the cavity 1 through the readout waveguide. The second term denotes the interaction between the cavity 1 and the readout waveugide, which yields a noise input $B_{\mathrm{in}}$ to the cavity, as shown laterly. Eq. \eqref{HamiltonianI} descibes couplings between the cavities and the reservoirs with strength $Y_{\mathrm{i}}$ and $Z_{\mathrm{i}}$, respectively. Notably, the position-dependent phase $e^{ikx_{m}}, $ ($m=1,2,3,4$) with a wave vector $k$ is introduced by the multi-point couplings.  

\subsection{Langevin Equations}
For the purpose of sensing, we analysis how the output varies when the perturbation $\varepsilon$ acts on the sensor, which can be done with the quantum Langevin equation. Before we proceed, we assume that the coupling points are equally spaced; i.e., $d=x_2-x_1=x_3-x_2=x_4-x_3$, and for simplicity, we let $x_1=0$.  Also, the linear dispersion relation holds in the dissipative reservoir; i.e., $\omega_{d,k}=v_g k$ with $v_g$ being the group velocity  [\cite{Zhu2021, Shen2009, Zhu2019}]. With the above assumptions, the equations of motion for two cavities take the form
\begin{subequations}
\begin{eqnarray}
&&\dot{\tilde{a}}_1[t]=F_{11}[\varepsilon]\tilde{a}_1[t]-2\pi|Z_1|^2\tilde{a}_1[t-\tau]e^{i\omega_L\tau}+F_{12}[\varepsilon]\tilde{a}_{2}[t]-\tilde{M}^{\mathrm{in}}_1[t],\label{EOM1}\\
&&\dot{\tilde{a}}_2[t]=F_{22}[\varepsilon]\tilde{a}_2[t]-2\pi|Z_2|^2\tilde{a}_2[t-\tau]e^{i\omega_L\tau}+F_{21}[\varepsilon]\tilde{a}_{1}[t]-F_{21}^{\mathrm{dir}}[t]-\tilde{M}^{\mathrm{in}}_2[t],\label{EOM2}
\end{eqnarray}
\end{subequations}
where 
\begin{subequations}
\begin{eqnarray}
&&F_{11}[\varepsilon]=i\omega_L-iH_{11}[\varepsilon]+\pi|Y_1|^2-2\pi|Z_1|^2-\frac{\kappa}{2},\\
&&F_{22}[\varepsilon]=i\omega_L-iH_{22}[\varepsilon]+\pi|Y_2|^2-2\pi|Z_2|^2,\\
&&F_{12}[\varepsilon]=-iH_{12}[\varepsilon]+\pi Y_1Y_2^\ast,\\
&&F_{21}[\varepsilon]=-iH_{21}[\varepsilon]+\pi Y_2Y_1^\ast,\\
&&F_{21}^{\mathrm{dir}}[t]=2\pi Z_2Z_1^\ast e^{i\omega_L\tau}\big(\tilde{a}_1[t-\tau]+2\tilde{a}_1[t-2\tau]e^{i\omega_L\tau}+\tilde{a}_1[t-3\tau]e^{2i\omega_L\tau}\big),\label{DirectionalCoupling}\\
&&\tilde{M}^{\mathrm{in}}_1[t]=i\sqrt{\kappa}\big(\beta+\tilde{B}_{\mathrm{in}}[t]\big)-i\sqrt{2\pi}Y_1 \tilde{C}_{\mathrm{in}}^\dagger[t]-i\sqrt{2\pi}Z_1\big(\tilde{D}_{\mathrm{in}}[t]+\tilde{D}_{\mathrm{in}}[t-\tau]e^{i\omega_L\tau}\big),\\
&&\tilde{M}^{\mathrm{in}}_2[t]=i\sqrt{2\pi}Y_2\tilde{C}_{\mathrm{in}}^\dagger[t]-i\sqrt{2\pi}Z_2e^{2i\omega_L\tau}\big(\tilde{D}_{\mathrm{in}}[t-2\tau]+\tilde{D}_{\mathrm{in}}[t-3\tau]e^{i\omega_L\tau}\big),
\end{eqnarray}
\end{subequations}
with $\tau=d/v_g$ being the time delay between the two neighboring points. Here, $\tilde{a}_{\mathrm{i}}[t]=a_{\mathrm{i}}[t]e^{i\omega_L t}$ denotes the slowly-varying operator. Also, 
\begin{subequations}
\begin{eqnarray}
&&\tilde{B}_{\mathrm{in}}[t]=B_{\mathrm{in}}[t]e^{i\omega_L t}=\frac{1}{\sqrt{2\pi}}\int\mathrm{d}k \ b_k[0]e^{-i(\omega_{b,k}-\omega_L)t},\\
&&\tilde{C}_{\mathrm{in}}^\dagger [t]=C_{\mathrm{in}}^\dagger [t]e^{i\omega_L t}=\frac{1}{\sqrt{2\pi}}\int\mathrm{d}k \ c_k^\dagger[0]e^{i(\omega_{c,k}+\omega_L)t},\\
&&\tilde{D}_{\mathrm{in}}[t]=D_{\mathrm{in}}[t]e^{i\omega_L t}=\frac{1}{\sqrt{2\pi}}\int\mathrm{d}k \ d_k[0]e^{-i(\omega_{d,k}-\omega_L)t},
\end{eqnarray}
\end{subequations}
are the  inputs for the readout waveguide, gain and dissipative reservoirs, respectively. In addition, the input-output relation for the field in the readout waveguide is given by 
\begin{equation}
\tilde{B}_{\mathrm{out}}[t]=\big(\tilde{B}_{\mathrm{in}}[t]+\beta\big)-i\sqrt{\kappa}\tilde{a}_1[t], \label{Inpu-Output}
\end{equation}
where
\begin{equation}
\tilde{B}_{\mathrm{out}}[t]=B_{\mathrm{out}}[t]e^{i\omega_L t}=\frac{1}{\sqrt{2\pi}}\int\mathrm{d}k\ b_k[t_1]e^{-i\omega_{b,k}(t-t_1)}e^{i\omega_L t}	
\end{equation}
is the output field in the waveguide at a final time $t_1$. 
  
Using Fourier transformation, the time-delayed differential equations \eqref{EOM1} and \eqref{EOM2} can be solved as  
 \begin{equation}
 \left(
\begin{array}{l}
\bar{a}_1[\omega;\varepsilon]\\
\bar{a}_2[\omega;\varepsilon]
\end{array}
\right)=\left((\omega_L+\omega) I-H[\varepsilon]-\pi i G_Y +2\pi i D_Z +\frac{i\tilde{\kappa}}{2}\right)^{-1}\bar{M}_{\mathrm{in}}[\omega]=\frac{\chi[\omega;\varepsilon]}{i\kappa}\bar{M}_{\mathrm{in}}[\omega],\label{GiantSolution}
 \end{equation}
with $\tilde{\kappa}=\kappa\left(\begin{array}{ll} 1 & 0 \\ 0 & 0 \end{array}\right)$, 
\begin{equation}
G_Y=\left(
 \begin{array}{ll}
  |Y_1|^2 & Y_1Y_2^\ast \\ 
  Y_1^\ast Y_2 & |Y_2|^2 
 \end{array}\right)
 =\left(\begin{array}{l}
Y_1\\
Y_2
\end{array}\right)
\left(\begin{array}{ll}
Y_1^\ast & Y_2^\ast\\
\end{array}\right)=YY^\dagger,\label{GY}
\end{equation}
\begin{equation}
D_Z=\left(\begin{array}{ll}
|Z_1|^2 (1+e^{i(\omega_L-\omega)\tau})& 0 \\
Z_2Z_1^\ast (e^{i(\omega_L-\omega)\tau}+2e^{i(2\omega_L-\omega)\tau}+e^{i(3\omega_L-\omega)\tau})& |Z_2|^2(1+e^{i(\omega_L-\omega)\tau})
\end{array}\right),\label{GiantDZ}	
\end{equation}
and 
\begin{eqnarray}
\bar{M}_{\mathrm{in}}[\omega]=&&
\left(\sqrt{\kappa}
\big(\begin{array}{ll}
2\pi\beta\delta[\omega]+\bar{B}_{\mathrm{in}}[\omega]\\
0	
\end{array}\big)
+\sqrt{2\pi}\big(\begin{array}{ll}
Y_1\\
Y_2
\end{array}\big)\bar{C}_{\mathrm{in}}^\dagger[\omega]+\right.\nonumber\\
&&\left.\sqrt{2\pi}\big(\begin{array}{ll}
Z_1 (1+e^{i(\omega_L-\omega)\tau})\\
Z_2 (e^{i(2\omega_L-\omega)\tau}+e^{i(3\omega_L-\omega)\tau})
\end{array}\big)\bar{D}_{\mathrm{in}}[\omega]
\right).\label{GiantInput}
\end{eqnarray}
Here, $I$ denotes a $2\times2$ identity matrix and $\chi[\omega;\varepsilon]$ is the dimensionless state transfer matrix. Operators with a bar $\bar{\cdot}$ denote the Fourier transformation of the corresponding operators in frequency domain. The diagonal terms in the gain matrix \eqref{GY} and dissipative matrix \eqref{GiantDZ} describe decays to the reservoirs, while the off-diagonal terms represent indirect couplings between the two cavities induced by the shared reservoir.  Different from Eq. \eqref{GY}, the non-Hermitianity of Eq. \eqref{GiantDZ} shows that the arrangement of the giant cavities can induce a non-reciprocal coupling $a_2\rightarrow a_1$ which results from the delayed coupling term $F^{\mathrm{dir}}_{21}[t]$ \eqref{DirectionalCoupling}. Notably, such a directional coupling is an inherent property in our proposal. Another change induced by the arrangement lies in the last term of the input matrix \eqref{GiantInput}, where exponents describe delayed inputs. Similarly, the input-output relation \eqref{Inpu-Output} in the frequency domain reads
\begin{eqnarray}
\bar{B}_{\mathrm{out}}[\omega]=&&(1-\chi_{11}[\omega;\varepsilon])(\bar{B}_{\mathrm{in}}[\omega]+2\pi\beta\delta[\omega])-\sqrt{\frac{2\pi}{\kappa}}\bar{C}_{\mathrm{in}}^\dagger[\omega]\left(\chi_{11}[\omega;\varepsilon]Y_1+\chi_{12}[\omega;\varepsilon]Y_2\right)-\nonumber\\
&&\sqrt{\frac{2\pi}{\kappa}}\bar{D}_{\mathrm{in}}[\omega]\left(\chi_{11}[\omega;\varepsilon]Z_1(1+e^{i(\omega_L-\omega)\tau})+\chi_{12}[\omega;\varepsilon]Z_2(e^{i(2\omega_L-\omega)\tau}+e^{i(3\omega_L-\omega)\tau})\right).
\label{InOutRe}
\end{eqnarray}

We have given a description of our sensor in the Heisenberg picture. From the above derivation, we can investigate how the unknown parameter affect the output of the detecting field. Different from the exsiting sensors, the dynamics of our sensor involve time-delay terms which would improve the performance of the sensor. 
\section{Performance evaluation of the sensor}
\subsection{Homodyne Detection}
As we have introduced, our sensor employs homodyne detection to evaluate the perturbation, where the photon current of the output field 
\begin{equation}
 I(t)=\sqrt{\frac{\kappa}{2}} (e^{i\varphi}B_{\mathrm{out}}(t)+H.c.),\label{PhotonCurrent}
\end{equation}
is measured. All the information of $\varepsilon$ is contained in the real part of $e^{i\varphi}B_{\mathrm{out}}[t]$. Note that the current is measured in a steady-state of the system such that we can evaluate the response of the system to the purterbation at the zero frequency; i.e., $\omega=0$. Also, for small $\varepsilon$, the expectation value of the output is assumed to be in a linear response to $\varepsilon$ [\cite{Lau2018}]; i.e., 
\begin{equation}
	\langle\bar{B}_{\mathrm{out}}[0]\rangle_\varepsilon=\langle\bar{B}_{\mathrm{out}}[0]\rangle_0+\lambda\varepsilon,\label{LinearResponse}
\end{equation}
where $\langle\cdot\rangle_z$ denotes taking expectations at $\varepsilon=z$. Using this relation, the response coefficient $\lambda$ reads 
\begin{equation}
	\lambda=\lim_{\varepsilon\rightarrow0}\frac{\langle\bar{B}_{\mathrm{out}}[0]\rangle_\varepsilon-\langle\bar{B}_{\mathrm{out}}[0]\rangle_0}{\varepsilon}
	=-2\pi\beta\delta[0]\frac{\mathrm{d}\chi_{11}[0;\varepsilon]}{\mathrm{d}\varepsilon}|_{\varepsilon=0}=\frac{2\pi i \beta\delta[0]}{\kappa}\left(\tilde{\chi}V\tilde{\chi}\right)_{11}\label{Lambda}
\end{equation}
whose phase determines the angle in Eq. (\ref{PhotonCurrent}); i.e., $\varphi=-\arg \lambda$.

\subsection{Signal, Noise, and Signal-to-noise Ratio per Photon}
To evaluate the performance of the sensor, we further define a measurement operator $m[\omega]$ as the windowed Fourier transformation of current $I[t]$, i.e., 
\begin{equation}
m[\omega]=\frac{1}{\sqrt{T}}\int_{-T/2}^{T/2}\mathrm{d}t\ I[t] e^{-i\omega t},\label{Measurement} 	
\end{equation}
where the segment $T$ should be much greater than $1/\kappa$ such that the sensor can reach the steady states during the measurement window. Under this condition, the integral limits can be extended to $\pm\infty$. Notably, this definition of $m[\omega]$ makes it have a unit of $A /\sqrt{\mathrm{Hz}}$ [\cite{Clerk2010}]. 

The power associated with signal can be defined as the square of difference of measurement operator $m[0]$ between the perturbed and unperturbed cases, i.e.,
\begin{equation}
S=\left(\langle m[0]\rangle_{\varepsilon}-\langle m[0]\rangle_0\right)^2=\frac{2\kappa\varepsilon^2}{T}|\lambda|^2.
\label{Signal}
\end{equation}
In addition, the total average photon number induced by the classical input can be calculated as 
\begin{equation}
n_{\mathrm{tot}}=\sum_{\mathrm{i}=1}^2	\langle \bar{a}_{\mathrm{i}}^\dagger[0;\varepsilon]\rangle_0 \langle \bar{a}_{\mathrm{i}}[0;\varepsilon]\rangle_0=\frac{|2\pi\delta[0]\beta|^2}{\kappa}\left(\tilde{\chi}^\dagger\tilde{\chi}\right)_{11}, \label{PhotonNumber}
\end{equation}
where the mean-field approximation [\cite{Huang2010, Qu2013}] has been used. With this definitaion, the signal per photon can be expressed as 
\begin{equation}
\frac{S}{n_{\mathrm{tot}}}=\frac{2\varepsilon^2}{T}\frac{|\big(\tilde{\chi}V\tilde{\chi}\big)_{11}|^2}{\big(\tilde{\chi}^\dagger\tilde{\chi}\big)_{11}},
\label{SperPhoton}	
\end{equation}
where we let $\tilde{\chi}=\chi[0;0]$ for brevity.

Similarly, the power of the output noise is defined as the fluctuation of the measurement operator $m[0]$ in the unperturbed case; i.e.
 \begin{eqnarray}
 N&&=\langle m^2[0]\rangle_0-\langle m[0]\rangle_0^2\nonumber\\
 &&\frac{\kappa}{2T}\left(1+|\tilde{\chi}_{11}|^2-(\tilde{\chi}_{11}+\tilde{\chi}_{11}^\ast)+\frac{2\pi}{\kappa}(\tilde{\chi} G_Y\tilde{\chi}^\dagger )_{11}+\frac{2\pi}{\kappa}(1+e^{i\omega_L \tau})^2(\tilde{\chi}\tilde{Z}\tilde{Z}^\dagger\tilde{\chi}^\dagger)_{11}\right)\nonumber\\
 &&=\frac{\kappa}{2T}\left(1+2\Xi\cdot\theta[\Xi]+\frac{4\pi}{\kappa}(1+\cos(\omega_L\tau))(1+e^{i\omega_L\tau})|Z_1\tilde{\chi}_{11}+Z_2\tilde{\chi}_{12}e^{2i\omega_L\tau}|^2\right),
\label{ReducedNoise}
 \end{eqnarray}
 where $\tilde{Z}=\left(\begin{array}{ll} Z_1 & Z_2 e^{2i\omega_L \tau} \end{array}\right)^T$, $\Xi=|\tilde{\chi}_{11}-1|^2-1$, and $\theta[\cdot]$ is the Heaviside step function introduced by the semi-defined positivity of the matrix $\tilde{\chi}G_Y \tilde{\chi}^\dagger$. In the derivation, we have assumed that both reservoirs and the waveguide are initially prepared in the vacuum states.  Note that the output noise \eqref{ReducedNoise} is complex due to the exponent $e^{i\omega_L \tau}$, which is in contrast to  Refs. [\cite{Lau2018, Bao2021}]. However, one can define its real part $\mathrm{Re} (N)$ as the measured noise. The constant part is the so-called shot noise [\cite{Lau2018}], which describes the minimum noise of the sensor. The second term  denotes the reflective gain resulting from the gain reservoir. When the sensor has a reflective gain; i.e., $|\tilde{\chi}_{11}-1|>1$, the output noise must be greater than the simple shot noise. Or equivalently speaking, a linear amplification for signal also amplifies the noise. And the third term results from the dissipative noise of the dissipative reservoir. 

Combing Eqs. \eqref{SperPhoton} and \eqref{ReducedNoise}, one can obtain the signal-to-noise ratio (SNR) per photon 
\begin{eqnarray}
\frac{S}{Nn_{\mathrm{tot}}}=\frac{4\varepsilon^2}{\kappa}\frac{|(\tilde{\chi}V\tilde{\chi})_{11}|^2}{\left(1+2\Xi\cdot\theta(\Xi)+\frac{4\pi}{\kappa}(1+\cos(\omega_L\tau))(1+e^{i\omega_L\tau})|Z_1\tilde{\chi}_{11}+Z_2\tilde{\chi}_{12}e^{2i\omega_L\tau}|^2\right)\left(\tilde{\chi}^\dagger\tilde{\chi}\right)_{11}}
\label{Signal-Noise}
\end{eqnarray}
which is the sensitivity of the sensor. Notably, the state transfer matrix $\tilde{\chi}$ is now independent in the perturbation $\varepsilon$, which means that the SNR has a purely parabolic response to the changes of $\varepsilon$ for a determined $\tilde{\chi}$.
\subsection{Corresponding results for the sensor composed of two small cavities}
For comparison, we also consider the sensor made up of two small cavities who couple to the dissipative reservoir in a single point way. This is a standard model of two-mode quantum sensors [\cite{Bao2021, Lau2018}] as a benchmark. In this case, the second line in interaction Hamiltonian Eq. \eqref{HamiltonianI} is rewritten as 
\begin{equation}
H_{I-D}^S=\sum_{\mathrm{i}=1}^2\int\mathrm{d}k \left(Z_{\mathrm{i}}	a_{\mathrm{i}}^\dagger d_k+H.c.\right).
\end{equation}
 This induces a modification on Eq. \eqref{GiantDZ}
 \begin{equation}
D_Z^S=\frac{1}{2}\left(\begin{array}{ll}
|Z_1|^2 & Z_1Z_2^\ast\\
Z_2Z_1^\ast & |Z_2|^2
\end{array}\right)
=\frac{1}{2}\left(\begin{array}{l}
Z_1\\
Z_2
\end{array}\right)
\left(\begin{array}{ll}
Z_1^\ast & Z_2^\ast\\
\end{array}\right)=\frac{1}{2}ZZ^\dagger
\label{SmallDZ}
 \end{equation}
and Eq. \eqref{GiantInput}  
\begin{equation}
\bar{M}_{\mathrm{in}}^S[0]=
\left(\sqrt{\kappa}
\big(\begin{array}{ll}
2\pi\beta\delta[0]+\bar{B}_{\mathrm{in}}[0]\\
0	
\end{array}\big)
+\sqrt{2\pi}\big(\begin{array}{ll}
Y_1\\
Y_2
\end{array}\big)\bar{C}_{\mathrm{in}}^\dagger[0]
+\sqrt{2\pi}\big(\begin{array}{ll}
Z_1\\
Z_2
\end{array}\big)\bar{D}_{\mathrm{in}}[0]
\right),\label{SmallInput}	
\end{equation}
and the gain matrix $G_Y$ \eqref{GY} remains the same. Hereafter, we use superscript $S$ to label the corresponding quantities of the sensor composed of small cavities. 

An interesting fact is that in this case the third term in Eq. \eqref{ReducedNoise} then reduces to $\frac{4\pi}{\kappa}|Z_1\tilde{\chi}_{11}^S+Z_2\tilde{\chi}_{12}^S|^2$ which is an unavoidable and untunable noise. However, in our proposal, one can adjust $\omega_L$ or $\tau$ to eliminate the dissipative noise such that the output noise $N$ can remain at a lower level.   

\section{Numerical comparison of giant vs small sensors}
 To numerically evaluate the performance of the sensor, we set the Hamiltonians $H^f[0]$ and $V$ as
\begin{equation}
H^f=\left(\begin{array}{ll}
\omega_1 & J \\
J & \omega_2
\end{array}\right)\ \ \mathrm{and} \ \ \ \ V=\left(\begin{array}{ll}
1 & 1\\
1 & 1
\end{array}\right),
\label{SpecificH}
\end{equation} 
which describes a common linear coupled-cavity system. For simplicity, we consider that both $Y_{\mathrm{i}}$ and $Z_{\mathrm{i}}$ are real. With these specific matrices, one can easily rewrite the state transfer matrix as 
\begin{eqnarray}
&&\tilde{\chi}=i\kappa\left(
\big(\begin{array}{ll}
\Delta-\frac{i\Gamma}{2}+\frac{i\kappa}{2} & -J-\frac{i\Gamma}{2} \\
-J-\frac{i\Gamma}{2} & \Delta+\Delta_{12}-\frac{i\Gamma}{2}
\end{array}\big)+\gamma(1+e^{i(\Delta\tau+\phi)})(\begin{array}{ll}
1 & 0 \\
e^{i(\Delta\tau+\phi)}+e^{2i(\Delta\tau+\phi)}& 1
\end{array}\big)
\right)^{-1},\\
&&\tilde{\chi}^S=i\kappa\left(
\big(\begin{array}{ll}
\Delta-\frac{i\Gamma}{2}+\frac{i\kappa}{2} & -J-\frac{i\Gamma}{2} \\
-J-\frac{i\Gamma}{2} & \Delta+\Delta_{12}-\frac{i\Gamma}{2}
\end{array}\big)+\frac{\gamma}{2}(\begin{array}{ll}
1 & 1 \\
1 & 1
\end{array}\big)
\right)^{-1},
\end{eqnarray}
where $\Delta=\omega_L-\omega_1$ and $\Delta_{12}=\omega_1-\omega_2$ are detunings, $\Gamma=2\pi Y_1^2=2\pi Y_2^2$ and  $\gamma=2\pi Z_1^2=2\pi Z_2^2$ denote the decay rates of the cavities to the reservoirs , and $\phi=\omega_1\tau$ is a fixed phase.
\begin{figure}[h!]
\begin{center}
\includegraphics[width=17.5cm]{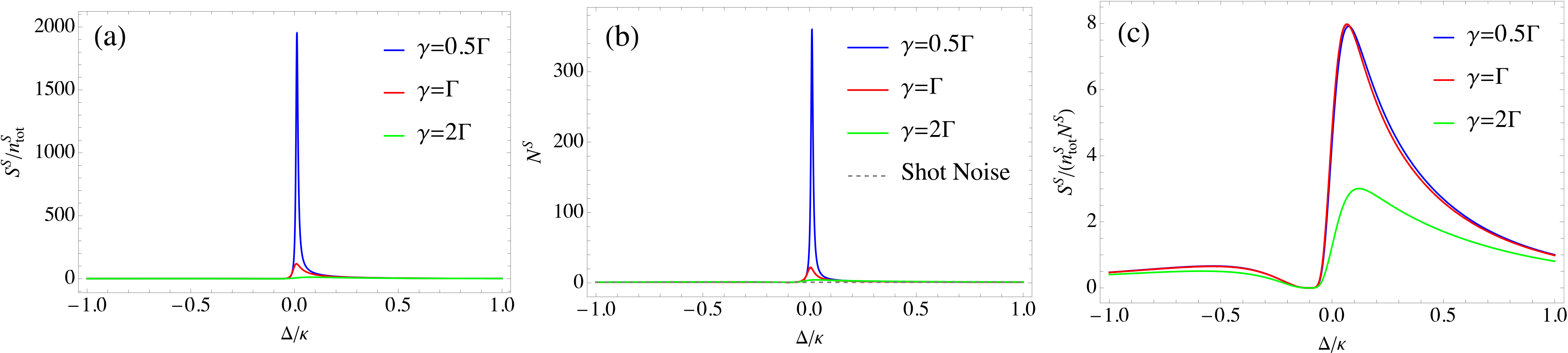}
\end{center}
\caption{(Color online) Spectra of relative signal per photon, noise, and SNR. Parameters in plotting are $\Delta_{12}=0$ and $J=\Gamma=0.1\kappa$. (a) Spectra of the relative signal per photon. The signal reaches the maximum at the resonant point, and decreases as the loss $\gamma$ increases; (b) Spectra of the relative noise. Similar to (a), the noise reaches the maximum at the resonant point, and decreases as the loss $\gamma$ increases; (c) Spectra of SNR. SNR does not reach its maximum value at the resonant point.}
\label{Small}
\end{figure}

\begin{figure}[h!]
\begin{center}
\includegraphics[width=16.8cm]{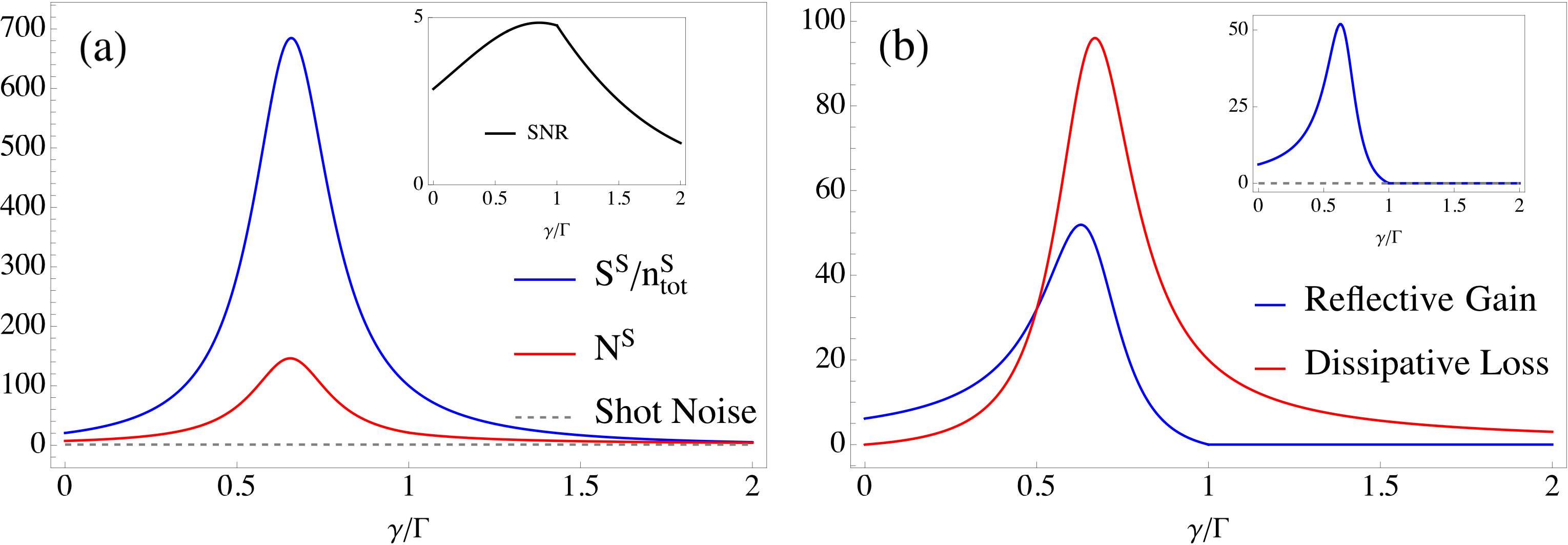}
\end{center}
\caption{(Color online) Relative signal per photon and noise as functions of $\gamma$ at the resonant point. Parameters in plotting are: $\Delta=\Delta_{12}=0$, $J=\Gamma=0.1\kappa$. (a) Both signal (blue line) and noise (red line) experience a process of first increase and then decrease, and reach their maximum value at $\gamma\simeq0.65\Gamma$, but SNR (black line in inset) reaches the maximum value at $\gamma\simeq0.85\Gamma$. Besides, the noise $N^S$ is always greater than the shot noise (Gray dotted line). (b) The reflective gain (Blue line, the second term in Eq. \eqref{ReducedNoise}) and the dissipative loss (Red line, the third term in Eq. \eqref{ReducedNoise}) as functions of $\gamma$ with the replacement of matrix $D^S_Z$. Inset: the amplified curves of reflective gain. One can clearly see that it is zero when $\gamma>\Gamma$, because the Heaviside function in Eq. \eqref{ReducedNoise} cutoff the parts $|\tilde{\chi}_{11}^S-1|^2-1<0$.} 
\label{Fig3}
\end{figure}

For numerical simulations, we set parameters $\Delta_{12}=0$ and $J=\Gamma=0.1\kappa$ which describes a good cavity in the weak coupling regime [\cite{Weis2010}]. We first plot the spectra of relative signal per photon, noise, and SNR of the sensor made up of two small cavities, as shown in Fig. \ref{Small}. A general conclusion is that both signal and noise reach the maximum value at the resonant point, as shown in Fig. \ref{Small} (a) and (b), but does not the SNR, as shown in Fig. \ref{Small} (c); Besides, in Fig. \ref{Small} (c), SNR with balanced gain and loss, i.e., $\gamma=\Gamma$ (red line), seems to be greater than other cases, but it is not indeed. To clarify this, we re-plot above curves as the functions of loss $\gamma$ at the resonance point in Fig. \ref{Fig3}. Hereafter, we only consider the cases at the resonant point. As the loss $\gamma$ increases, both signal and noise gradually increase until reach their maximum values at $\gamma\simeq 0.65\Gamma$ and then decrease, shown as the blue and red lines in Fig. \ref{Fig3} (a). Intuitively, it seems like that SNR also should reach its maximum value at this point, but regrettably,  the maximum value of SNR occurs at $\gamma\simeq0.85\Gamma$, as shown in the inset of Fig. \ref{Fig3} (a). Besides, in the inset of Fig. \ref{Fig3} (a), an interesting fact is that, a sudden change of SNR occurs at the balanced gain and loss $\gamma=\Gamma$. This is because $|\tilde{\chi}_{11}-1|^2<1$ when $\gamma>\Gamma$, such that the reflective gain in Eq. \eqref{ReducedNoise} is cutoff by the Heaviside function, which means it does not make sense, as shown as the blue line in Fig.\ref{Fig3} (b).  In this regime, the total noise only includes the shot noise and the dissipative noise. Another point to be clarified is, why the signal and noise experience a process of first increase and then decrease rather than monotonously increase, as the loss $\gamma$ increases and the gain $\Gamma$ is fixed. This is because the state transfer matrix $\tilde{\chi}^S$ is also a function of $\gamma$, where the element $|\tilde{\chi}^S_{11}|$ dominating the reflective gain has two extreme points $\gamma\simeq0.65\Gamma$ and  $\gamma=\Gamma$, while the element $|\tilde{\chi}^S_{12}|$ dominating the dissipative loss has only one extreme point $\gamma\simeq0.65\Gamma$. This property makes the signal and noise, as the compound function of $\gamma$ and $\tilde{\chi}$, experience the process of first increase and then decrease. 

\begin{figure}[h!]
\begin{center}
\includegraphics[width=16.8cm]{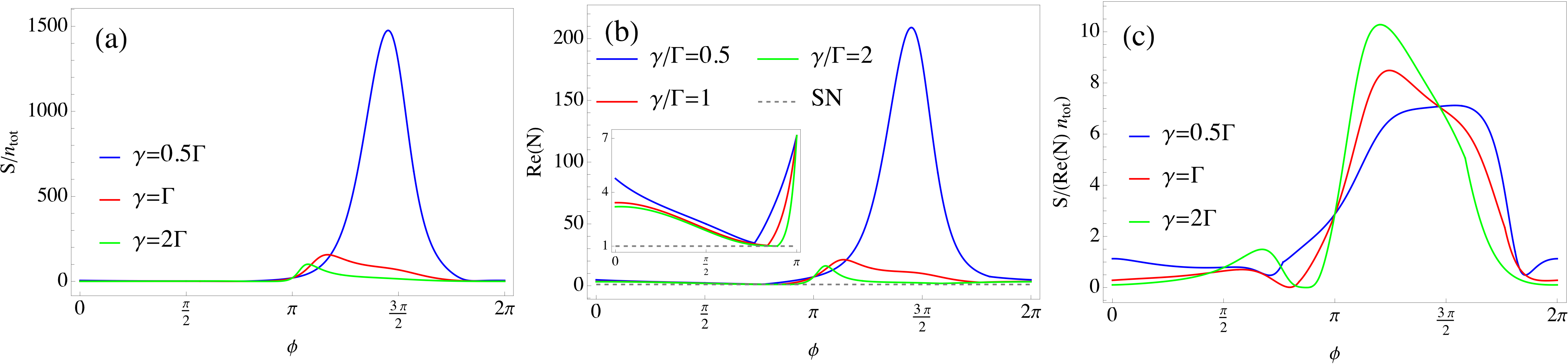}
\end{center}
\caption{(Color online) Relative signal per photon, noise and SNR as the functions of  $\phi$ at the resonant point. Parameters in plotting are: $\Delta=\Delta_{12}=0$, $J=\Gamma=0.1\kappa$. (a) The relative signal as the function of $\phi$. As the loss $\gamma$ increases, the maximum value of the signal decreases; (b) The relative noise as the function of $\phi$. The term "SN" is the abbreviation for Shot noise. Similar to (a), the noise also increases as the loss $\gamma$ increases, and it is always greater than the shot noise. However, at some certain $\phi$, the noise can remain at the shot noise level,  e.g., $N\simeq1.18$ at $\phi=0.76\pi$ when $\gamma=0.5\Gamma$, $N\simeq1.04$ at $\phi=0.84\pi$ when $\gamma=\Gamma$ and $N\simeq1.01$ at $\phi=0.89\pi$ when $\gamma=2\Gamma$. (c) The relative SNR as the function of $\phi$. Similar to (a) and (b), SNR also increases as the loss $\gamma$ increases.} 
\label{Fig4}
\end{figure}

\begin{figure}[h!]
\begin{center}
\includegraphics[width=16.8cm]{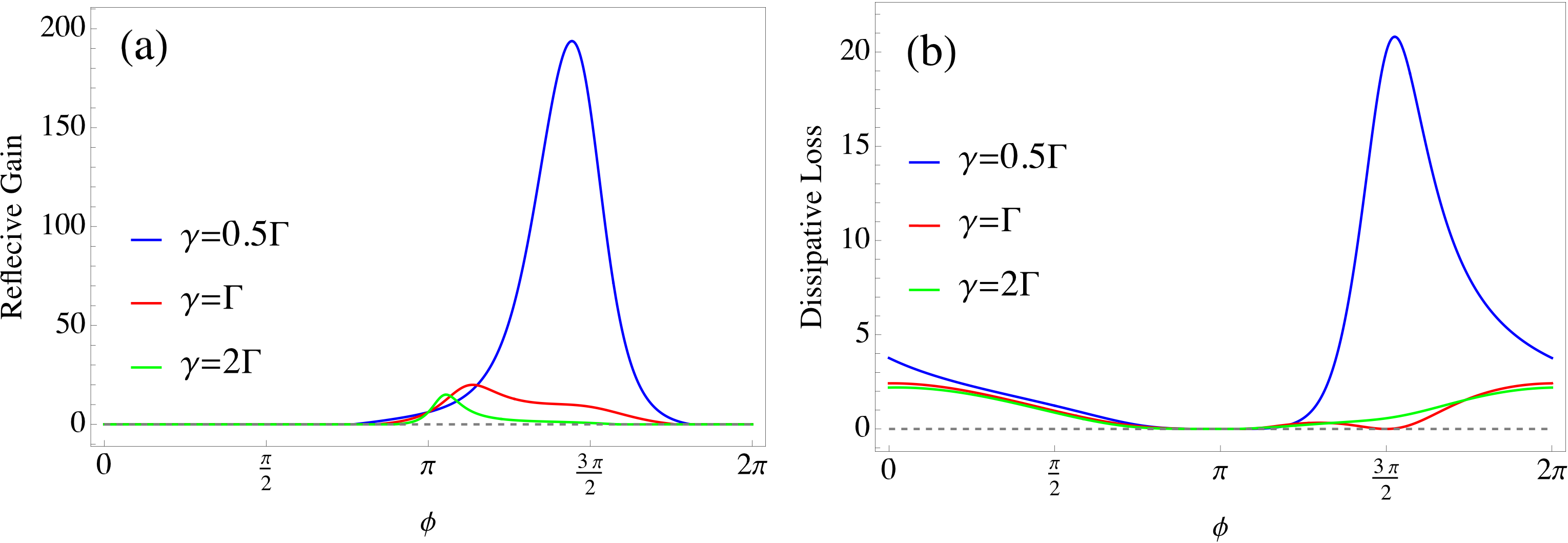}
\end{center}
\caption{(Color online) Reflective gain (a) and dissipative loss (b) as the functions of $\phi$. Parameters in plotting are: $\Delta=\Delta_{12}=0$, $J=\Gamma=0.1\kappa$. Both reflective gain and dissipative loss can be zero at some certain $\phi$ but they cannon be zero simultaneously, which is the reason why the noise is always greater than the shot noise in Fig. \ref{Fig4} (b).} 
\label{Fig5}
\end{figure}

With the previous results, we now turn to the sensor made up of two giant cavities. The dissipative matrix $D_Z$ additionally introduces a degree of freedom of fixed phase $\phi$ ($\Delta\tau$ is zero at the resonant point). Using the same parameters, the relative signal per photon, noise and SNR as the functions of phase $\phi$ are plotted in Fig. \ref{Fig4}. Both signal and noise also experience a process of first increase and then decrease as the phase $\phi$ increases, as shown in Fig.\ref{Fig4} (a) and (b); An interesting point is that, thanks to the phase $\phi$, the noise can remain at the shot noise level, e.g., $N\simeq1.18$ at $\phi=0.76\pi$ when $\gamma=0.5\Gamma$ (Blue line), $N\simeq1.04$ at $\phi=0.84\pi$ when $\gamma=\Gamma$ (Red line) and $N\simeq1.01$ at $\phi=0.89\pi$ when $\gamma=2\Gamma$ (Green line), which are about one order of magnitude smaller than $N^S$, as shown as the inset in Fig. \ref{Fig4} (b). This is our third central result; In Fig. \ref{Fig4} (c), It seems like that SNR increases as the loss $\gamma$ increases during $\phi\in[\pi,2\pi]$, but indeed, SNR reaches its maximum value at $\gamma\simeq\Gamma$. \footnote{We have simulated SNR with $\gamma\in\{0.25\Gamma, 0.5\Gamma, \Gamma, 2\Gamma, 4\Gamma, 8\Gamma, 16\Gamma\}$ and found SNR is maximum at $\gamma=4\Gamma$. For the sake of keeping the picture simple and clear, we do not show other curves in Fig. \ref{Fig4} (c).}  As for the reflective gain and dissipative loss, although they can be zero by adjusting the phase $\phi$, they cannot be zero simultaneously, and this property also explains why the noise is always greater than the shot noise in Fig. \ref{Fig4} (b), as shown in Fig. \ref{Fig5}. 

\begin{figure}[h!]
\begin{center}
\includegraphics[width=16.8cm]{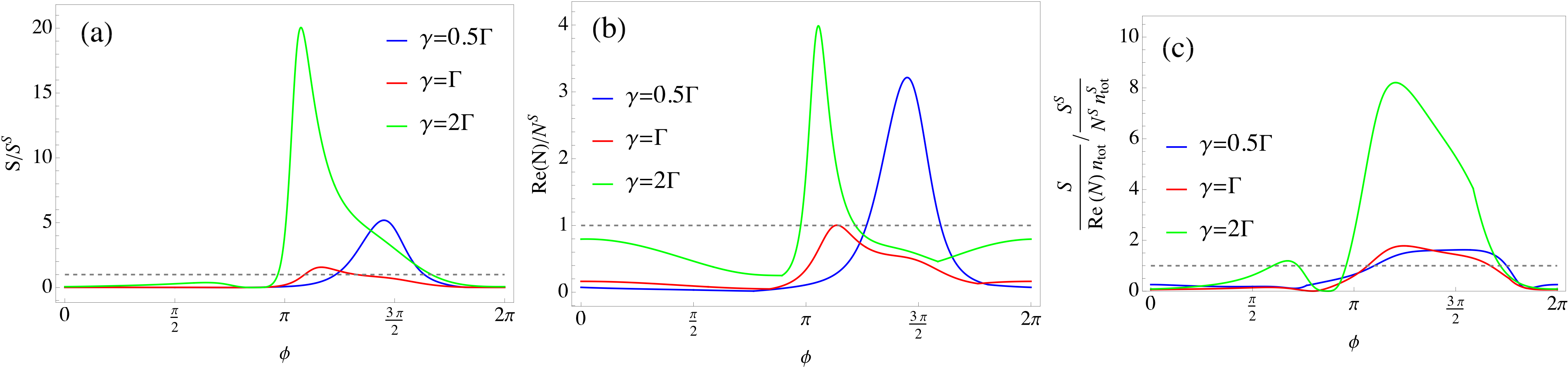}
\end{center}
\caption{(Color online) The ratios of signal (a), output noise (a) and SNR (c) between sensors made up of giant and small cavities. Parameters in plotting are: $\Delta=\Delta_{12}=0$ and $J=\Gamma=0.1\kappa$. (a) For some certain intervals, e.g., $[\pi, 1.75\pi]$, the signals $S$ are greater than $S^S$. Especially, when $\gamma=2\Gamma$, $S$ is about one order of magnitude enhanced compared to the sensor made up of small cavities. (b) By properly adjusting phase $\phi$, $\mathrm{Re} (N)$ can be lower than $N^S$. In particularly, when $\gamma=\Gamma$, $\mathrm{Re} (N)$ is smaller than $N^S$ globally. (c) With proper gain and loss, e.g., $\gamma=2\Gamma$, SNR is about one order magnitude greater than $S^S/(N^S n^S_{\mathrm{tot}})$. These results show that the giant-cavity structure is a powerful resource in designing quantum sensors.} 
\label{Fig6}
\end{figure}

To make a clear comparison with the sensors made up of small cavities, we also plot Fig. \ref{Fig6}. As Fig.\ref{Fig6} (a) shows, the signal $S$ can be about one order of magnitude greater than $S^S$, especially when $\gamma=2\Gamma$ (green line). An interesting point is that, $\mathrm{Re}(N)$ is always smaller than $N^S$ in the entire interval $[0, 2\pi]$ when $\gamma=\Gamma$, as shown as the red line in Fig. \ref{Fig6} (b). This means that our proposal can effectively decrease the output noise by adjusting parameters $\phi$, compared to Ref. [\cite{Lau2018}]. More importantly, although both signal $S$ and output noise $\mathrm{Re} (N)$ are enhanced in the interval $[\pi,1.75\pi]$ when $\gamma=2\Gamma$ (green lines in Fig. \ref{Fig6} (a) and (b), respectively), SNR is much greater than other conditions of $\gamma$, as shown as the green line in Fig. \ref{Fig6} (c). These  results show that the giant-cavity structure is a powerful resource in designing quantum sensors. 

 \section{Conclusion and Future Works}
In summary, we proposed a quantum sensor consisting of two giant cavities. By coupling cavities to a dissipative reservoir at multiple points, a non-reciprocal interaction can be engineered between the cavities and the common reservoir, which requires no non-linear elements. Compared to the standard two-mode quantum sensor [\cite{Lau2018}], the output noise can remain at the shot noise level, which is reduced by about one order of magnitude. Thus in turn increases the signal-to-noise ratio per photon by about one order of magnitude. These results show that the giant cavity-based sensor can effectively improve the sensing precision. 

A future direction is to take the non-Markovian effect induced by large time delay $\tau$ into account [\cite{Guo2017, Andersson2019, Zhu2021}]. Since we only consider the case at resonant point, the non-Markovian effect related to $\Delta\tau$ is negelcted. However, this degree of freedom plays important roles in the deep non-Markovian regime $\tau\gg1/\kappa$ [\cite{Guo2017}], e.g., it induces a non-exponential decay [\cite{Andersson2019}] and a multi-peak excitation spectrum [\cite{Zhu2021}]. How these non-Markovian effects affect the sensing performance is an open question to be explored in the future, especially when an coherent feedback is applied to control the system [\cite{Xue2012, Xue2016, Xue2017a, Xue2017b, Xue2020}]. 
\section*{Acknowledgments}
We thank Lei Du, Wenlong Li, Qiuyuan Cai, and Sulin Feng for the fruitful discussion. This work is supported by the National Natural Science Foundation of China (NSFC) under Grants 61873162, 61973317. This work was also supported by the Open Research Project of the State Key Laboratory of Industrial Control Technology, Zhejiang University, China (No. ICT2022B47).

\bibliographystyle{Frontiers-Harvard} 
\bibliography{sensor-03-13}

\begin{thebibliography}{55}
\providecommand{\natexlab}[1]{#1}
\expandafter\ifx\csname urlstyle\endcsname\relax
  \providecommand{\doi}[1]{doi:\discretionary{}{}{}#1}\else
  \providecommand{\doi}{doi:\discretionary{}{}{}\begingroup
  \urlstyle{rm}\Url}\fi
\providecommand{\selectlanguage}[1]{\relax}
\providecommand{\bibAnnoteFile}[1]{%
  \IfFileExists{#1}{\begin{quotation}\noindent\textsc{Key:} #1\\
  \textsc{Annotation:}\ \input{#1}\end{quotation}}{}}
\providecommand{\bibAnnote}[2]{%
  \begin{quotation}\noindent\textsc{Key:} #1\\
  \textsc{Annotation:}\ #2\end{quotation}}

\bibitem[{Andersson et~al.(2019)Andersson, Suri, Guo, Aref, and
  Delsing}]{Andersson2019}
Andersson, G., Suri, B., Guo, L., Aref, T., and Delsing, P. (2019).
\newblock Non-exponential decay of a giant artificial atom.
\newblock \emph{Nat. Phys.} 15, 1123--1127.
\newblock \doi{10.1038/s41567-019-0605-6}
\bibAnnoteFile{Andersson2019}

\bibitem[{Bao et~al.(2021)Bao, Qi, Dong, and Nori}]{Bao2021}
Bao, L., Qi, B., Dong, D., and Nori, F. (2021).
\newblock Fundamental limits for reciprocal and nonreciprocal non-hermitian
  quantum sensing.
\newblock \emph{Phys. Rev. A} 103, 042418.
\newblock \doi{10.1103/PhysRevA.103.042418}
\bibAnnoteFile{Bao2021}

\bibitem[{Cai and Jia(2021)}]{Cai2021}
Cai, Q.~Y. and Jia, W.~Z. (2021).
\newblock Coherent single-photon scattering spectra for a giant-atom
  waveguide-qed system beyond the dipole approximation.
\newblock \emph{Phys. Rev. A} 104, 033710.
\newblock \doi{10.1103/PhysRevA.104.033710}
\bibAnnoteFile{Cai2021}

\bibitem[{Casimir(1945)}]{Casmir1945}
Casimir, H. B.~G. (1945).
\newblock On onsager's principle of microscopic reversibility.
\newblock \emph{Rev. Mod. Phys.} 17, 343--350.
\newblock \doi{10.1103/RevModPhys.17.343}
\bibAnnoteFile{Casmir1945}

\bibitem[{Clerk et~al.(2010)Clerk, Devoret, Girvin, Marquardt, and
  Schoelkopf}]{Clerk2010}
Clerk, A.~A., Devoret, M.~H., Girvin, S.~M., Marquardt, F., and Schoelkopf,
  R.~J. (2010).
\newblock Introduction to quantum noise, measurement, and amplification.
\newblock \emph{Rev. Mod. Phys.} 82, 1155--1208.
\newblock \doi{10.1103/RevModPhys.82.1155}
\bibAnnoteFile{Clerk2010}

\bibitem[{Delsing et~al.(2019)Delsing, Cleland, Schuetz, Kn\"{o}rzer, Giedke,
  Cirac et~al.}]{Delsing2019}
Delsing, P., Cleland, A.~N., Schuetz, M. J.~A., Kn\"{o}rzer, J., Giedke, G.,
  Cirac, J.~I., et~al. (2019).
\newblock The 2019 surface acoustic waves roadmap.
\newblock \emph{Journal of Physics D: Applied Physics} 52, 353001.
\newblock \doi{10.1088/1361-6463/ab1b04}
\bibAnnoteFile{Delsing2019}

\bibitem[{Du et~al.(2021{\natexlab{a}})Du, Cai, Wu, Wang, and Li}]{Du2021A}
Du, L., Cai, M.-R., Wu, J.-H., Wang, Z., and Li, Y. (2021{\natexlab{a}}).
\newblock Single-photon nonreciprocal excitation transfer with non-markovian
  retarded effects.
\newblock \emph{Phys. Rev. A} 103, 053701.
\newblock \doi{10.1103/PhysRevA.103.053701}
\bibAnnoteFile{Du2021A}

\bibitem[{Du and Li(2021)}]{Du2021B}
Du, L. and Li, Y. (2021).
\newblock Single-photon frequency conversion via a giant
  $\mathrm{\ensuremath{\Lambda}}$-type atom.
\newblock \emph{Phys. Rev. A} 104, 023712.
\newblock \doi{10.1103/PhysRevA.104.023712}
\bibAnnoteFile{Du2021B}

\bibitem[{Du et~al.(2021{\natexlab{b}})Du, Zhang, Wu, Kockum, and Li}]{Du2021C}
Du, L., Zhang, Y., Wu, J.-H., Kockum, A.~F., and Li, Y. (2021{\natexlab{b}}).
\newblock Giant atoms in synthetic frequency dimensions.
\newblock \emph{arXiv:} 2111.05584.
\bibAnnoteFile{Du2021C}

\bibitem[{Ekstr\"om et~al.(2019)Ekstr\"om, Aref, Ask, Andersson, Suri, Sanada
  et~al.}]{Ekstrom2019}
Ekstr\"om, M.~K., Aref, T., Ask, A., Andersson, G., Suri, B., Sanada, H.,
  et~al. (2019).
\newblock Towards phonon routing: controlling propagating acoustic waves in the
  quantum regime.
\newblock \emph{New J. Phys.} 21, 123013.
\newblock \doi{10.1088/1367-2630/ab5ca5}
\bibAnnoteFile{Ekstrom2019}

\bibitem[{Guo et~al.(2017)Guo, Grimsmo, Kockum, Pletyukhov, and
  Johansson}]{Guo2017}
Guo, L., Grimsmo, A., Kockum, A.~F., Pletyukhov, M., and Johansson, G. (2017).
\newblock Giant acoustic atom: A single quantum system with a deterministic
  time delay.
\newblock \emph{Phys. Rev. A} 95, 053821.
\newblock \doi{10.1103/PhysRevA.95.053821}
\bibAnnoteFile{Guo2017}

\bibitem[{Guo et~al.(2020)Guo, Kockum, Marquardt, and Johansson}]{Guo2020}
Guo, L., Kockum, A.~F., Marquardt, F., and Johansson, G. (2020).
\newblock Oscillating bound states for a giant atom.
\newblock \emph{Phys. Rev. Research} 2, 043014.
\newblock \doi{10.1103/PhysRevResearch.2.043014}
\bibAnnoteFile{Guo2020}

\bibitem[{Huang and Agarwal(2010)}]{Huang2010}
Huang, S. and Agarwal, G.~S. (2010).
\newblock Reactive-coupling-induced normal mode splittings in microdisk
  resonators coupled to waveguides.
\newblock \emph{Phys. Rev. A} 81, 053810.
\newblock \doi{10.1103/PhysRevA.81.053810}
\bibAnnoteFile{Huang2010}

\bibitem[{Kannan et~al.(2020)Kannan, Ruckriegel, Campbell, Kockum, Braumuller,
  Kim et~al.}]{Kannan2020}
Kannan, B., Ruckriegel, M.~J., Campbell, D.~L., Kockum, A.~F., Braumuller, J.,
  Kim, D.~K., et~al. (2020).
\newblock Waveguide quantum electrodynamics with superconducting artificial
  giant atoms.
\newblock \emph{Nature} 583, 775--779.
\newblock \doi{10.1038/s41586-020-2529-9}
\bibAnnoteFile{Kannan2020}

\bibitem[{Kockum(2021)}]{Kockum2020}
Kockum, A.~F. (2021).
\newblock Quantum optics with giant atoms---the first five years.
\newblock In \emph{International Symposium on Mathematics, Quantum Theory, and
  Cryptography} (Springer Singapore), 125--146
\bibAnnoteFile{Kockum2020}

\bibitem[{Kockum et~al.(2014)Kockum, Delsing, and Johansson}]{Kockum2014}
Kockum, A.~F., Delsing, P., and Johansson, G. (2014).
\newblock Designing frequency-dependent relaxation rates and lamb shifts for a
  giant artificial atom.
\newblock \emph{Phys. Rev. A} 90, 013837.
\newblock \doi{10.1103/PhysRevA.90.013837}
\bibAnnoteFile{Kockum2014}

\bibitem[{Kockum et~al.(2018)Kockum, Johansson, and Nori}]{Kockum2018}
Kockum, A.~F., Johansson, G., and Nori, F. (2018).
\newblock {Decoherence-free interaction between giant atoms in waveguide
  quantum electrodynamics}.
\newblock \emph{Phys. Rev. Lett.} 120, 140404.
\newblock \doi{10.1103/physrevlett.120.140404}
\bibAnnoteFile{Kockum2018}

\bibitem[{Konotop and Kuzmiak(2002)}]{Konotop2002}
Konotop, V.~V. and Kuzmiak, V. (2002).
\newblock Nonreciprocal frequency doubler of electromagnetic waves based on a
  photonic crystal.
\newblock \emph{Phys. Rev. B} 66, 235208.
\newblock \doi{10.1103/PhysRevB.66.235208}
\bibAnnoteFile{Konotop2002}

\bibitem[{Lau and Clerk(2018)}]{Lau2018}
Lau, H.-K. and Clerk, A.~A. (2018).
\newblock Fundamental limits and non-reciprocal approaches in non-hermitian
  quantum sensing.
\newblock \emph{Nature Communications} 9, 4320.
\newblock \doi{10.1038/s41467-018-06477-7}
\bibAnnoteFile{Lau2018}

\bibitem[{Li et~al.(2018{\natexlab{a}})Li, Bilek, Hoff, Madsen, Forstner,
  Prakash et~al.}]{Li2018B}
Li, B.-B., Bilek, J., Hoff, U.~B., Madsen, L.~S., Forstner, S., Prakash, V.,
  et~al. (2018{\natexlab{a}}).
\newblock Quantum enhanced optomechanical magnetometry.
\newblock \emph{Optica} 5, 850--856.
\newblock \doi{10.1364/OPTICA.5.000850}
\bibAnnoteFile{Li2018B}

\bibitem[{Li et~al.(2018{\natexlab{b}})Li, Bulla, Prakash, Forstner,
  Dehghan-Manshadi, Rubinsztein-Dunlop et~al.}]{Li2018A}
Li, B.-B., Bulla, D., Prakash, V., Forstner, S., Dehghan-Manshadi, A.,
  Rubinsztein-Dunlop, H., et~al. (2018{\natexlab{b}}).
\newblock Scalable high-sensitivity optomechanical magnetometers on a chip.
\newblock \emph{APL Photonics} 3, 120806.
\newblock \doi{10.1063/1.5055029}
\bibAnnoteFile{Li2018A}

\bibitem[{Manenti et~al.(2017)Manenti, Kockum, Patterson, Behrle, Rahamim,
  Tancredi et~al.}]{Manenti2017}
Manenti, R., Kockum, A.~F., Patterson, A., Behrle, T., Rahamim, J., Tancredi,
  G., et~al. (2017).
\newblock Circuit quantum acoustodynamics with surface acoustic waves.
\newblock \emph{Nat. Comm.} 8, 975.
\newblock \doi{10.1038/s41467-017-01063-9}
\bibAnnoteFile{Manenti2017}

\bibitem[{Marks(2014)}]{Marks2014}
Marks, P. (2014).
\newblock Quantum positioning system steps in when gps fails.
\newblock \emph{New Scientist} 222, 19.
\newblock \doi{https://doi.org/10.1016/S0262-4079(14)60955-6}
\bibAnnoteFile{Marks2014}

\bibitem[{McDonald and Clerk(2020)}]{McDonald2020}
McDonald, A. and Clerk, A.~A. (2020).
\newblock Exponentially-enhanced quantum sensing with non-hermitian lattice
  dynamics.
\newblock \emph{Nature Communications} 11, 5382.
\newblock \doi{10.1038/s41467-020-19090-4}
\bibAnnoteFile{McDonald2020}

\bibitem[{Peng et~al.(2014)Peng, Ozdemir, Lei, Monifi, Gianfreda, Long
  et~al.}]{Peng2014B}
Peng, B., Ozdemir, S.~K., Lei, F., Monifi, F., Gianfreda, M., Long, G.~L.,
  et~al. (2014).
\newblock Parity–time-symmetric whispering-gallery microcavities.
\newblock \emph{Nature Physics} 10, 394--398.
\newblock \doi{10.1038/nphys4323}
\bibAnnoteFile{Peng2014B}

\bibitem[{Peng et~al.(2016)Peng, {\"O}zdemir, Liertzer, Chen, Kramer, Y{\i}lmaz
  et~al.}]{Peng2016}
Peng, B., {\"O}zdemir, {\c S}.~K., Liertzer, M., Chen, W., Kramer, J.,
  Y{\i}lmaz, H., et~al. (2016).
\newblock Chiral modes and directional lasing at exceptional points.
\newblock \emph{Proceedings of the National Academy of Sciences} 113,
  6845--6850.
\newblock \doi{10.1073/pnas.1603318113}
\bibAnnoteFile{Peng2016}

\bibitem[{Phare et~al.(2015)Phare, Daniel~Lee, Cardenas, and
  Lipson}]{Phare2015}
Phare, C.~T., Daniel~Lee, Y.-H., Cardenas, J., and Lipson, M. (2015).
\newblock Graphene electro-optic modulator with 30 $\mathrm{GHz}$ bandwidth.
\newblock \emph{Nature Photonics} 9, 511--514.
\newblock \doi{10.1038/nphoton.2015.122}
\bibAnnoteFile{Phare2015}

\bibitem[{Potton(2004)}]{Potton2004}
Potton, R.~J. (2004).
\newblock Reciprocity in optics.
\newblock \emph{Reports on Progress in Physics} 67, 717--754.
\newblock \doi{10.1088/0034-4885/67/5/r03}
\bibAnnoteFile{Potton2004}

\bibitem[{Qu and Agarwal(2013)}]{Qu2013}
Qu, K. and Agarwal, G.~S. (2013).
\newblock Phonon-mediated electromagnetically induced absorption in hybrid
  opto-electromechanical systems.
\newblock \emph{Phys. Rev. A} 87, 031802.
\newblock \doi{10.1103/PhysRevA.87.031802}
\bibAnnoteFile{Qu2013}

\bibitem[{Scalora et~al.(1994)Scalora, Dowling, Bowden, and
  Bloemer}]{Scalora1994}
Scalora, M., Dowling, J.~P., Bowden, C.~M., and Bloemer, M.~J. (1994).
\newblock The photonic band edge optical diode.
\newblock \emph{Journal of Applied Physics} 76, 2023--2026.
\newblock \doi{10.1063/1.358512}
\bibAnnoteFile{Scalora1994}

\bibitem[{Schnabel et~al.(2010)Schnabel, Mavalvala, McClelland, and
  Lam}]{Schnabel2010}
Schnabel, R., Mavalvala, N., McClelland, D.~E., and Lam, P.~K. (2010).
\newblock Quantum metrology for gravitational wave astronomy.
\newblock \emph{Nature Communications} 1, 121.
\newblock \doi{10.1038/ncomms1122}
\bibAnnoteFile{Schnabel2010}

\bibitem[{Schuetz et~al.(2015)Schuetz, Kessler, Giedke, Vandersypen, Lukin, and
  Cirac}]{Schuetz2015}
Schuetz, M. J.~A., Kessler, E.~M., Giedke, G., Vandersypen, L. M.~K., Lukin,
  M.~D., and Cirac, J.~I. (2015).
\newblock Universal quantum transducers based on surface acoustic waves.
\newblock \emph{Phys. Rev. X} 5, 031031.
\newblock \doi{10.1103/PhysRevX.5.031031}
\bibAnnoteFile{Schuetz2015}

\bibitem[{Shen and Fan(2009)}]{Shen2009}
Shen, J.-T. and Fan, S. (2009).
\newblock Theory of single-photon transport in a single-mode waveguide. i.
  coupling to a cavity containing a two-level atom.
\newblock \emph{Phys. Rev. A} 79, 023837.
\newblock \doi{10.1103/PhysRevA.79.023837}
\bibAnnoteFile{Shen2009}

\bibitem[{Shi et~al.(2014)Shi, Kong, Wang, Kong, Zhao, Liu et~al.}]{Shi2014}
Shi, F., Kong, X., Wang, P., Kong, F., Zhao, N., Liu, R.-B., et~al. (2014).
\newblock Sensing and atomic-scale structure analysis of single nuclear-spin
  clusters in diamond.
\newblock \emph{Nature Physics} 10, 21--25.
\newblock \doi{10.1038/nphys2814}
\bibAnnoteFile{Shi2014}

\bibitem[{Sánchez-Burillo et~al.(2012)Sánchez-Burillo, Duch,
  Gómez-Gardeñes, and Zueco}]{Sanchez-Burillo2012}
Sánchez-Burillo, E., Duch, J., Gómez-Gardeñes, J., and Zueco, D. (2012).
\newblock Quantum navigation and ranking in complex networks.
\newblock \emph{Scientific Reports} 2, 605.
\newblock \doi{10.1038/srep00605}
\bibAnnoteFile{Sanchez-Burillo2012}

\bibitem[{Sounas and Alu(2017)}]{Sounas2017}
Sounas, D.~L. and Alu, A. (2017).
\newblock Non-reciprocal photonics based on time modulation.
\newblock \emph{Nature Photonics} 11, 774--783.
\newblock \doi{10.1038/s41566-017-0051-x}
\bibAnnoteFile{Sounas2017}

\bibitem[{Stray et~al.(2022)Stray, Lamb, Kaushik, Vovrosh, Rodgers, Winch
  et~al.}]{Stray2022}
Stray, B., Lamb, A., Kaushik, A., Vovrosh, J., Rodgers, A., Winch, J., et~al.
  (2022).
\newblock Quantum sensing for gravity cartography.
\newblock \emph{Nature} 602, 590--594.
\newblock \doi{10.1038/s41586-021-04315-3}
\bibAnnoteFile{Stray2022}

\bibitem[{Tocci et~al.(1995)Tocci, Bloemer, Scalora, Dowling, and
  Bowden}]{Tocci1995}
Tocci, M.~D., Bloemer, M.~J., Scalora, M., Dowling, J.~P., and Bowden, C.~M.
  (1995).
\newblock Thin‐film nonlinear optical diode.
\newblock \emph{Applied Physics Letters} 66, 2324--2326.
\newblock \doi{10.1063/1.113970}
\bibAnnoteFile{Tocci1995}

\bibitem[{Vadiraj et~al.(2021)Vadiraj, Ask, McConkey, Nsanzineza, Chang, Kockum
  et~al.}]{Vadiraj2021}
Vadiraj, A.~M., Ask, A., McConkey, T.~G., Nsanzineza, I., Chang, C. W.~S.,
  Kockum, A.~F., et~al. (2021).
\newblock Engineering the level structure of a giant artificial atom in
  waveguide quantum electrodynamics.
\newblock \emph{Phys. Rev. A} 103, 023710.
\newblock \doi{10.1103/PhysRevA.103.023710}
\bibAnnoteFile{Vadiraj2021}

\bibitem[{Vollmer et~al.(2008)Vollmer, Arnold, and Keng}]{Vollmer2008}
Vollmer, F., Arnold, S., and Keng, D. (2008).
\newblock Single virus detection from the reactive shift of a
  whispering-gallery mode.
\newblock \emph{Proceedings of the National Academy of Sciences} 105,
  20701--20704.
\newblock \doi{10.1073/pnas.0808988106}
\bibAnnoteFile{Vollmer2008}

\bibitem[{Wang et~al.(2013)Wang, Zhou, Guo, Zhang, Evers, and Zhu}]{Wang2013}
Wang, D.-W., Zhou, H.-T., Guo, M.-J., Zhang, J.-X., Evers, J., and Zhu, S.-Y.
  (2013).
\newblock Optical diode made from a moving photonic crystal.
\newblock \emph{Phys. Rev. Lett.} 110, 093901.
\newblock \doi{10.1103/PhysRevLett.110.093901}
\bibAnnoteFile{Wang2013}

\bibitem[{Weis et~al.(2010)Weis, Rivière, Deleglise, Gavartin, Arcizet,
  Schliesser et~al.}]{Weis2010}
Weis, S., Rivière, R., Deleglise, S., Gavartin, E., Arcizet, O., Schliesser,
  A., et~al. (2010).
\newblock Optomechanically induced transparency.
\newblock \emph{Science} 330, 1520--1523.
\newblock \doi{10.1126/science.1195596}
\bibAnnoteFile{Weis2010}

\bibitem[{Xu et~al.(2005)Xu, Schmidt, Pradhan, and Lipson}]{Xu2005}
Xu, Q., Schmidt, B., Pradhan, S., and Lipson, M. (2005).
\newblock Micrometre-scale silicon electro-optic modulator.
\newblock \emph{Nature} 435, 325--327.
\newblock \doi{10.1038/nature03569}
\bibAnnoteFile{Xu2005}

\bibitem[{Xu et~al.(2018)Xu, Chen, Zhao, Li, Lu, and Yang}]{Xu2018}
Xu, X., Chen, W., Zhao, G., Li, Y., Lu, C., and Yang, L. (2018).
\newblock Wireless whispering-gallery-mode sensor for thermal sensing and
  aerial mapping.
\newblock \emph{Light: Science and Applications} 7, 62.
\newblock \doi{10.1038/s41377-018-0063-4}
\bibAnnoteFile{Xu2018}

\bibitem[{Xue et~al.(2017{\natexlab{a}})Xue, Hush, and Petersen}]{Xue2017b}
Xue, S., Hush, M.~R., and Petersen, I.~R. (2017{\natexlab{a}}).
\newblock Feedback tracking control of non-markovian quantum systems.
\newblock \emph{IEEE Transactions on Control Systems Technology} 25,
  1552--1563.
\newblock \doi{10.1109/TCST.2016.2614834}
\bibAnnoteFile{Xue2017b}

\bibitem[{Xue et~al.(2020)Xue, Nguyen, James, Shabani, Ugrinovskii, and
  Petersen}]{Xue2020}
Xue, S., Nguyen, T., James, M.~R., Shabani, A., Ugrinovskii, V., and Petersen,
  I.~R. (2020).
\newblock Modeling for non-markovian quantum systems.
\newblock \emph{IEEE Transactions on Control Systems Technology} 28,
  2564--2571.
\newblock \doi{10.1109/TCST.2019.2935421}
\bibAnnoteFile{Xue2020}

\bibitem[{Xue and Petersen(2016)}]{Xue2016}
Xue, S. and Petersen, I.~R. (2016).
\newblock Realizing the dynamics of a non-markovian quantum system by markovian
  coupled oscillators: a green’s function-based root locus approach.
\newblock \emph{Quantum Information Processing} 15, 1001--1018.
\newblock \doi{10.1007/s11128-015-1196-5}
\bibAnnoteFile{Xue2016}

\bibitem[{Xue et~al.(2017{\natexlab{b}})Xue, Wu, Hush, and Tarn}]{Xue2017a}
Xue, S., Wu, R., Hush, M.~R., and Tarn, T.-J. (2017{\natexlab{b}}).
\newblock Non-markovian coherent feedback control of quantum dot systems.
\newblock \emph{Quantum Science and Technology} 2, 014002.
\newblock \doi{10.1088/2058-9565/aa6125}
\bibAnnoteFile{Xue2017a}

\bibitem[{Xue et~al.(2012)Xue, Wu, Zhang, Zhang, Li, and Tarn}]{Xue2012}
Xue, S., Wu, R., Zhang, W., Zhang, J., Li, C., and Tarn, T.-J. (2012).
\newblock Decoherence suppression via non-markovian coherent feedback control.
\newblock \emph{Phys. Rev. A} 86, 052304.
\newblock \doi{10.1103/PhysRevA.86.052304}
\bibAnnoteFile{Xue2012}

\bibitem[{Yu et~al.(2016)Yu, Janousek, Sheridan, McAuslan, Rubinsztein-Dunlop,
  Lam et~al.}]{Yu2016}
Yu, C., Janousek, J., Sheridan, E., McAuslan, D.~L., Rubinsztein-Dunlop, H.,
  Lam, P.~K., et~al. (2016).
\newblock Optomechanical magnetometry with a macroscopic resonator.
\newblock \emph{Phys. Rev. Applied} 5, 044007.
\newblock \doi{10.1103/PhysRevApplied.5.044007}
\bibAnnoteFile{Yu2016}

\bibitem[{Zhao et~al.(2011)Zhao, Hu, Ho, Wan, and Liu}]{Zhao2011}
Zhao, N., Hu, J.-L., Ho, S.-W., Wan, J. T.~K., and Liu, R.~B. (2011).
\newblock Atomic-scale magnetometry of distant nuclear spin clusters via
  nitrogen-vacancy spin in diamond.
\newblock \emph{Nature Nanotechnology} 6, 242--246.
\newblock \doi{10.1038/nnano.2011.22}
\bibAnnoteFile{Zhao2011}

\bibitem[{Zhu et~al.(2010)Zhu, Ozdemir, Xiao, Li, He, Chen et~al.}]{Zhu2010}
Zhu, J., Ozdemir, S.~K., Xiao, Y.-F., Li, L., He, L., Chen, D.-R., et~al.
  (2010).
\newblock On-chip single nanoparticle detection and sizing by mode splitting in
  an ultrahigh-q microresonator.
\newblock \emph{Nature Photonics} 4, 46--49.
\newblock \doi{10.1038/nphoton.2009.237}
\bibAnnoteFile{Zhu2010}

\bibitem[{Zhu et~al.(2021)Zhu, Wu, and Xue}]{Zhu2021}
Zhu, Y., Wu, R., and Xue, S. (2021).
\newblock Spatial non-locality induced non-markovian eit in a single giant
  atom.
\newblock \emph{arXiv:} 2106.05020.
\bibAnnoteFile{Zhu2021}

\bibitem[{Zhu and Jia(2019)}]{Zhu2019}
Zhu, Y.~T. and Jia, W.~Z. (2019).
\newblock Single-photon quantum router in the microwave regime utilizing double
  superconducting resonators with tunable coupling.
\newblock \emph{Phys. Rev. A} 99, 063815.
\newblock \doi{10.1103/PhysRevA.99.063815}
\bibAnnoteFile{Zhu2019}

\bibitem[{Zhukovsky and Smirnov(2011)}]{Zhukovsky2011}
Zhukovsky, S.~V. and Smirnov, A.~G. (2011).
\newblock All-optical diode action in asymmetric nonlinear photonic multilayers
  with perfect transmission resonances.
\newblock \emph{Phys. Rev. A} 83, 023818.
\newblock \doi{10.1103/PhysRevA.83.023818}
\bibAnnoteFile{Zhukovsky2011}

\end{thebibliography}
\end{document}